\documentclass[useAMS,usenatbib,a4paper]{mn2e_mod}
\usepackage{epsfig,amsmath,amssymb,amsfonts,natbib,color,url}  

\usepackage{amsmath,amssymb,amsfonts,natbib}
\usepackage {graphicx,color}
\usepackage {txfonts }
\usepackage {natbib  }
\usepackage {float   }
\usepackage {enumerate}
\usepackage {subfigure  }

\newcommand{\be}{\begin{equation}}
\newcommand{\ee}{\end{equation}}
\newcommand \beq {\begin{equation}}
\newcommand \eeq {\end{equation}}

\newcommand \rhosolid  {\rho_{\rm solid}}
\newcommand \rhog  {\rho_{\rm gas}}
\newcommand \rhobulk  {\rho_{\rm bulk}}
\newcommand \rhocore  {\rho_{\rm core}}

\newcommand \RB {R_{\rm Bondi}}
\newcommand \RH {R_{\rm Hill}}

\newcommand \cg {c_{\rm gas}}

\newcommand \K {{\rm\,K}}  
\def\s{{\rm\,s}} 
\def\yr{{\rm\,yr}}

\def\cm{{\rm\,cm}} 
\def\m{{\rm\,m}} 
\def\km{{\rm\,km}} 
 
\def\gm{{\rm\,g}} 
\def\g{{\rm\,g}} 
 
\def\AU{{\rm AU}} 
 
\def\eV{{\rm eV}} 

\title[IN-SITU PLANET FORMATION]{The Minimum-Mass Extrasolar Nebula: \\
{\it In-Situ} Formation of Close-In Super-Earths}
\author[E.\,Chiang \&\,G.\,Laughlin]{Eugene Chiang$^{1}$\footnotemark[1] and\,Gregory Laughlin$^{2}$\footnotemark[1]\\
 \\
  $^1$Departments of Astronomy and of Earth and Planetary Science, University of California, Berkeley, Hearst Field Annex B-20, Berkeley CA 94720-3411, USA\\  $^2$Department of Astronomy and Astrophysics, UCO/Lick Observatory, University of California, Santa Cruz, Santa Cruz, CA 95064, USA\\
}

\date{Submitted: \today}

\begin{document}
\maketitle
\begin{abstract}
  Close-in super-Earths, with radii $R \approx 2$--$5R_\oplus$ and orbital
  periods $P < 100$ d, orbit more than half, and perhaps nearly all Sun-like
  stars in the universe.  We use this omnipresent population to
  construct the minimum-mass extrasolar nebula (MMEN), the
  circumstellar disk of solar-composition solids and gas from which
  such planets formed, if they formed near their current locations and
  did not migrate. In a series of back-of-the-envelope calculations,
  we demonstrate how {\it in-situ} formation in the MMEN is fast,
  efficient, and can reproduce many of the observed properties of
  close-in super-Earths, including their gas-to-rock
  fractions. Testable predictions are discussed.
\end{abstract}

\label{firstpage}
\footnotetext[1]{e-mail: \rm{\url{echiang@astro.berkeley.edu}}, \rm{\url{laugh@ucolick.org}} }

\section{INTRODUCTION}\label{sec_intro}

The Solar System has provided, and
continues to provide, the {\it de facto} template for most discussions of
planet formation.  But with a multitude of extrasolar worlds now
known, with masses approaching those of our terrestrial planets, we
can ask whether the Solar System's orbital architecture is
the norm --- or whether a new template is needed.

\subsection{Close-in super-Earths are ubiquitous} \label{sec:ubi}

The HARPS (High Accuracy Radial-Velocity Planet Searcher) project reported that $> 50$\% of chromospherically quiet, 
main-sequence dwarf stars in the Solar
neighborhood are accompanied by planets with masses $M \sin i \lesssim
30 \,M_{\oplus}$ and orbital periods $P < 100$ days --- hereafter
``close-in super-Earths'' \citep{mayor11}.
Such a startlingly large occurrence rate 
 appears consistent with the latest
results of the {\it Kepler} mission, as we now argue.
\citet{batalha12} used {\it Kepler} to survey
$N_{\ast,{\rm total}} = 1.56 \times 10^5$ stars --- almost all main-sequence solar-type dwarfs (G.~Marcy 2012, personal communication). Define:
\begin{itemize}
\item ${\cal F}_{\rm intrinsic}$ to be the fraction of {\it Kepler} dwarfs
that host at least one close-in super-Earth with radius $R > 2
R_\oplus$ and $P < 100$ days;
\item ${\cal F}_{\rm detect}$ to be the fraction of systems
with close-in super-Earths whose transit light curves achieve the
mandated 
detection threshold to be listed as candidates in the \citet{batalha12} catalog (i.e., among the subset of
{\it Kepler} dwarfs which actually do host close-in
super-Earths, $1- {\cal F}_{\rm detect}$ is the fraction
that are missing from the catalog);
\item ${\cal F}_{\rm transit} \approx 2.5$\% to be
the geometric probability of transit averaged over a distribution of
planets for which $dN / d\log P \propto P^{1/2}$ between $P= 7$ and 100 days \citep{youdin11}.\footnote{The best-fit slope of the period distribution of Youdin (2011) derives from only 
about half of the planet candidates found in the Batalha et al.~(2012) catalog, and moreover applies only up to $P = 50$ days. Fortunately, our calculation of the average ${\cal F}_{\rm transit}$ is not sensitive to the exact slope of the period distribution as long as it is fairly flat. For example, $dN/d\log P \propto P^0$ yields ${\cal F}_{\rm transit} = 3.0\%$.}
\end{itemize}
Batalha et al.~(2012) reported
that $N_{\ast} = 1.8 \times 10^3$ individual stars harbor $\sim$$2.3 \times 10^3$ planet candidates --- more than 80\% of which have radii
$R < 5 \,R_\oplus$. {\it In toto} we have
\begin{equation}
{\cal F}_{\rm intrinsic} {\cal F}_{\rm detect} {\cal F}_{\rm transit} N_{\ast,{\rm total}} = N_\ast
\end{equation}
or
\begin{equation}
{\cal F}_{\rm intrinsic} \approx 0.5
\left( \frac{1}{{\cal F}_{\rm detect}} \right) \left( \frac{N_\ast}{1.8\times 10^3} \right) \left( \frac{1.56 \times 10^5}{N_{\ast,{\rm total}}} \right) 
\left( \frac{0.025}{{\cal F}_{\rm transit}} \right) 
\end{equation}
which implies that ${\cal F}_{\rm intrinsic} \gtrsim 50$\%, in accord
with the occurrence rate calculated by Mayor et al.~(2011). \citet{figueira12}
performed a more careful consistency analysis between
the results of the HARPS and {\it Kepler} surveys, and found that the
two exoplanet populations can be reconciled --- and the large number
of multiple-transiting systems accounted for --- if planet-planet
mutual inclinations are less than $\sim$1 degree.\footnote{
Our equation (1) is strictly valid only for coplanar or single-planet
systems. A transit survey alone cannot determine $F_{\rm intrinsic}$
without assuming an inclination distribution for multi-planet systems
(see, e.g., \citealt{tremainedong12}).}

It seems clear that our Solar
System --- which contains no planet interior to Mercury's $P = 88$ day
orbit --- did not participate in a major if not the dominant mode of
planet formation in the Galaxy.

\subsection{Close-in super-Earths are a distinct population\\ with possible ties to giant planet satellites} \label{sec:sat}

\begin{figure*}
\centering
\includegraphics[width=\linewidth]{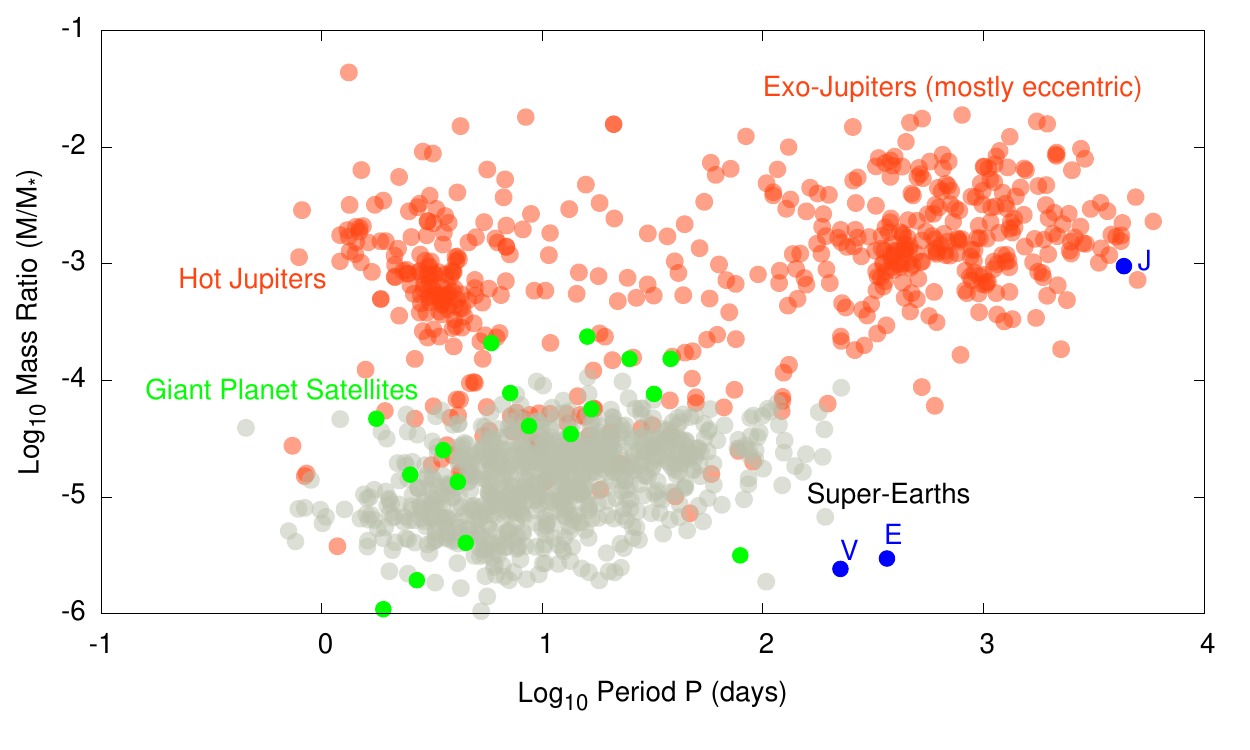}
\caption{A ``bird's eye view'' of published extrasolar planets.  We
  argue in this paper that hot Jupiters and super-Earths have distinct
  formation histories: hot Jupiters may have formed among the main
  population of exo-Jupiters at long orbital periods and migrated
  inward, whereas super-Earths formed {\it in situ}. Hot Jupiters are
  a fringe population as they are found orbiting only ${\cal F}_{\rm
    intrinsic} \approx 0.5$--1\% of Sun-like stars. By contrast,
  close-in super-Earths abound, with ${\cal F}_{\rm intrinsic} \gtrsim
  50$\%.  {\it Red circles:} Planets detected by the radial velocity
  method (either with or without photometric transits), taken from www.exoplanets.org on 07 Oct 2012 (see also \citealt{wright11}).
  Planet masses are $M \sin i$.  {\it Gray circles:} {\it Kepler}
  Objects of Interest (KOI) for which multiple KOIs are associated
  with a single target star. Radii, as reported in Bathalha et
  al.~(2012), are converted to masses by using
  $M/M_{\oplus}=(R/R_{\oplus})^{2.06}$, the best-fit power relation
  for planets in the Solar System (see \S\ref{wulith12} for alternative
  mass-radius relations).
  Only KOIs for which $R < 5 R_\oplus$ are plotted.
  {\it Green circles:} Regular
  satellites of the Jovian planets in our Solar System. Mass ratios
  are those of satellites to their host planets.  {\it Blue circles:}
  Jupiter, Earth and Venus.}
\label{fig1}
\end{figure*}

Aside from being commonplace, close-in super-Earths form a distinct
population in the space of $\log (M/M_\ast )$ and $\log P$, where
$M_\ast$ is the host primary mass. In Figure \ref{fig1}, super-Earths
are well separated from hot Jupiters, and are also distinct from
Jovian-mass planets with $P > 100$ days --- the latter having
generally eccentric orbits \citep{Zakamska11}.  Jupiter lies on the
fringe of the exo-Jupiter distribution, and its position on the edge
may be real. Arguably, the Doppler surveys have had more than adequate
time baseline and precision to uncover an abundance of true-Jupiter
analogs if they existed (Burt et al.~2013, in preparation, will present a comprehensive analysis of radial velocity measurements from the Keck-I iodine cell spanning the years 1998--2012).

The super-Earth population is characterized by (i) orbital periods
ranging from days to weeks, (ii) mass ratios $M/M_{\ast} \sim
10^{-4.5}$, and (iii) orbits that are co-planar to within a few
degrees (e.g., \citealt{fangmargot12}; \citealt{fabrycky12a}; \citealt{tremainedong12}). All these properties are reminiscent of the regular
satellite systems of Solar System giant planets. The fact that
Jupiter, Saturn, and Uranus\footnote{Presumably Neptune also once
  harbored a regular satellite system, which was destroyed when Triton
  was captured \citep{goldreich89}.}  all possess broadly similar
satellite systems indicates that the satellite formation process is
robust --- just as robust as the formation process for close-in
super-Earths, with which it may share more than a passing resemblance.

One difference between satellites and super-Earths is the propensity
of the former to be found in mean-motion resonances. Tidal
interactions with the host planet expand satellite orbits and drive
convergent migration into resonances \citep{murraydermott00}.  In the
case of close-in super-Earths, tidal changes to orbital semimajor axes
are typically less dramatic.  Nevertheless, tides may still shape
super-Earth orbits in observable ways: tidal dissipation can damp
orbital eccentricities and wedge near-resonant planets farther apart
(\citealt{LithWu12}; \citealt{BatMor12}). This process may explain the
observed excess of {\it Kepler} planet pairs just outside of resonance
(\citealt{lissrag11}; \citealt{fabrycky12a}).

\subsection{Migration vs.~{\it in-situ} accretion}

Giant planets with $P \ll 100$ days are commonly thought to
have formed at distances of several AUs from their host stars, and
then to have migrated to their current locations.  The difficulty of
forming hot Jupiters {\it in situ} is frequently acknowledged (e.g.,
Rafikov 2006). Migration has also been invoked to explain the close-in
orbits of lower-mass super-Earths (e.g., \citealt{alibert06};
\citealt{schlaufman09}; \citealt{lopez12}). Although much of the work
on orbital migration is well-motivated, we wonder how much stems from
an insistence (perhaps only implicit) on using the Solar System ---
with its obvious lack of close-in planets --- as a starting point for
studies of planet formation.\footnote{Migration may also have occurred
in our Solar System, but arguably not to the dramatic extents
imagined for close-in exoplanets. A classic and still viable scenario
for the migration of giant planets in our primordial planetesimal disk
posits that Jupiter migrated inward by a fraction of an AU
while Neptune migrated outward from $\sim$22 to 30 AU
(\citealt{fernip84}; \citealt{mal93}).}
To what extent is our view still
provincial, unduly colored by a naive interpretation of one system?

A popular mechanism for transporting planets inward is via
gravitational torques exerted by parent disks (see \citealt{kley12}
for a review). At the moment, disk-driven migration seems too poorly
understood to connect meaningfully with observations. Population
synthesis models that include prescriptions for disk-driven migration
fail to reproduce the observed statistics of planet occurrence at $P
\lesssim 50$ days (\citealt{idalin10}; \citealt{howard10};
\citealt{howard12}, and references therein). Theoretical uncertainties
include the effects of co-rotation resonances, disk thermodynamics
(which can even cause planets to migrate outward; see, e.g., section
2.2 in Kley \& Nelson 2012, and references therein), and the perennial
mystery of the source of viscosity in protoplanetary disks. It seems
premature to discuss how planets are transported in disks when we
cannot reliably say how disk material itself is transported.

Planets that migrate smoothly and converge on one another
  can trap themselves into mean-motion resonances, but most
  planetary systems do not exhibit such resonances.  Period ratios of
  multi-planet systems discovered by {\it Kepler} appear to be largely
  random, aside from a small preponderance of ratios just greater than
  3:2 and 2:1 that can be attributed to a modest amount of
  differential tidal decay (\citealt{lissrag11};
  \citealt{fabrycky12a}; \citealt{LithWu12}; \citealt{BatMor12}).
  Among more massive giant planets discovered by the radial velocity
  method, \citet{Wright11a} show that the occurrence rate of period
  commensurabilities is greater than random, but is still only about 1
  in 3 --- and in most cases resonant libration cannot be confirmed.
  Even for Gliese 876 b and c --- a pair of giant planets trapped
  deeply in a 2:1 resonance --- the amount by which these planets
  migrated differentially could have been as small as $\sim$7\%
  (\citealt{LeePeale02}; see also \citealt{Rivera10}).

Disk-driven migration is not the only means of migration.
Gravitational torques can be supplied instead by additional planetary
or even stellar companions.  Hot Jupiters whose orbit normals are
severely misaligned with the spin axes of their host stars seem most
naturally explained by dynamical instabilities that can deliver giant
planets onto high-eccentricity, high-inclination orbits.  Such orbits
can circularize by stellar tidal friction while retaining their large
inclinations (for a general overview of the dynamics, see
\citealt{wulith11}).  Although inward transport nicely explains
spin-misaligned hot Jupiters, we emphasize that hot Jupiters as a
whole represent but a small fraction of the entire close-in population
of planets --- i.e., hot Jupiters, and the need for large-distance
migration that they imply, may be the exception rather than the rule. The
occurrence rate of hot Jupiters is only about $\sim$1\% (e.g., Mayor
et al.~2011; Howard et al.~2012; from the Batalha et al.~2012 catalog
we estimate an occurrence rate of 0.5\%), in contrast to the
order-unity occurrence rate of close-in super-Earths
(\S\ref{sec:ubi}).  While migration of one kind or another may be
necessary to explain the fringe population represented by hot and even
warm Jupiters, the same may not be true for the majority of close-in
planetary systems --- particularly the ubiquitous super-Earths, lying
as they do in a distinct region of parameter space (Figure
\ref{fig1}).

Here we explore the possibility that long-distance orbital migration
does not play a major role in the genesis of close-in super-Earths ---
that such worlds formed instead {\it in situ} from circumstellar disks
of solids and gas extending interior to 0.5 AU. This idea is certainly
not new; it is nothing more than an extension of the
  classical formation scenario for our Solar System's terrestrial
  planets to distances inside Mercury's orbit.  Recent explorations
of {\it in-situ} accretion include those by Raymond et
al.~(\citeyear{Raymond08}; see their Table 1 and Figure 1),
\citet{Montgomery09}, and \citet{hansenmurray12}, and in many respects
our work parallels theirs. In a similar vein, \citet{ikomahori12}
calculated how close-in rocky cores accreted gas {\it in situ}, with
specific application to the gas-laden super-Earths orbiting
Kepler-11. Some of these papers still appealed to the
transport of solids from regions beyond 0.5 AU: Hansen \& Murray
(2012) invoked inward migration of planetesimals to furnish a massive
enough reservoir of raw rocks to form super-Earths, and Ikoma \& Hori
(2012) assumed that the progenitor cores of the Kepler-11 planets
migrated inward to their current locations. 

\subsection{Plan of this paper}

As we have argued above, close-in super-Earths are not anomalous ---
they are the norm, and it is our Solar System that is the exception to
the rule that the majority of main-sequence Sun-like stars harbor
planets with $R > 2 R_\oplus$ and $P < 100$ days. We compute from
scratch a ``minimum-mass extrasolar nebula'' (MMEN) using the
abundance of data from the {\it Kepler} mission
(\S\ref{sec_mmen}).\footnote{An earlier attempt to compute the MMEN
  using Doppler-detected planets is made by \citet{Kuchner04}.} How the
MMEN can spawn close-in super-Earths {\it in situ} is explored through
a series of back-of-the-envelope calculations that readers are
encouraged to reproduce and/or challenge (\S\ref{sec_oom}). {\it
  In-situ} formation leads to a number of predictions that seem ripe
for testing (\S\ref{sec:conc}).

\section{MINIMUM-MASS EXTRASOLAR NEBULA (MMEN)}\label{sec_mmen}

We construct the ``minimum-mass extrasolar nebula'' (MMEN): the
solar-metallicity disk of gas and solids out of which the super-Earths
uncovered by {\it Kepler} could have formed, if planet
formation were 100\% efficient and orbital migration were
negligible. We employ the $N = 1925$ planet candidates with radii $R <
5 R_{\oplus}$ and $P < 100$ days reported by Batalha at al.~(2012) for
the first 480 days of {\it Kepler} observations.

\begin{figure*}
\centering
\includegraphics[width=\linewidth]{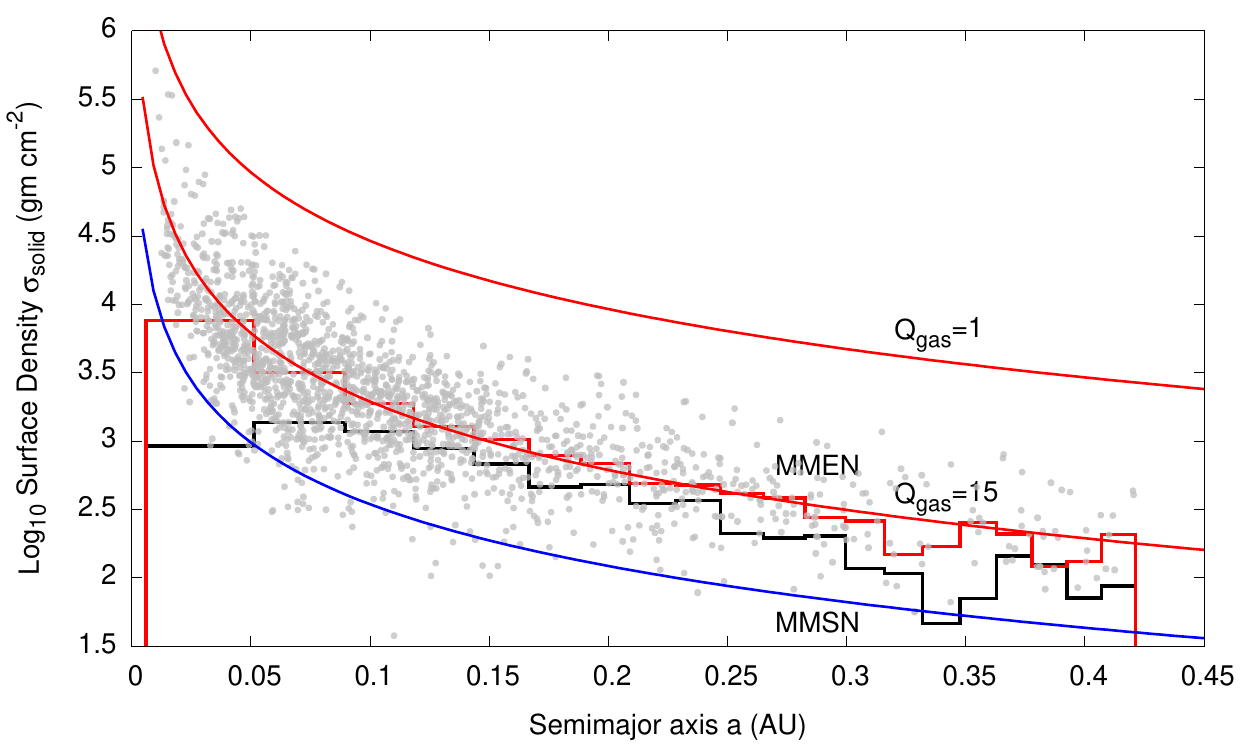}
\caption{The solid surface density profile of the ``minimum-mass
  extrasolar nebula'' (MMEN) constructed from {\it Kepler} data.  {\it
    Gray circles:} $\{\sigma_{{\rm solid},i}\} \equiv \{ M_i/2\pi
  a_i^2\}$ computed from {\it Kepler} planets with $R < 5 R_\oplus$
  and $P < 100$ d, assuming $M/M_\oplus = (R/R_\oplus)^{2.06}$ and
  solar-mass host stars.{\protect \footnotemark[3]}  The red histogram is the binned median
  of the gray points, and the solid red curve is the power law
  (equation \ref{eq_sigmad}) fitted to all red histogram bins except
  the last bin at $a < 0.05$ AU (see \S\ref{sec:alternate}). The black
  histogram is an alternate construction of the MMEN
  (\S\ref{sec:alternate}); we argue that it is lower than the red
  histogram at $a \gtrsim 0.05$ AU because of incompleteness in the
  {\it Kepler} catalog, and also lower at $a \lesssim 0.05$ AU because
  dust is typically absent at these distances in T Tauri disks and
  cannot seed planet formation. When enough water (oxygen), H, and He
  are added to solids to bring the entire MMEN up to solar composition
  (metallicity = 0.015), the resultant Toomre parameter $Q_{\rm gas}
  \approx 15$. If we allow for factors of 2--3 inefficiency in forming
  planets, then some primordial disks (those populating the upper
  envelope of gray points) approach the $Q_{\rm gas} = 1$ threshold
  for gravitational instability. For reference, $\sigma_{\rm solid}$
  for the minimum-mass Solar nebula (MMSN) --- or technically its
  extrapolation inward, since no planet is present interior to Mercury
  at $a = 0.4$ AU --- is plotted as a blue solid curve. As plotted, the MMEN is a
  factor of 5 times more massive than the MMSN at these distances, but this factor could be as low as $\sim$1 if other mass-radius relations are assumed (\S\ref{wulith12}).}
\label{fig2}
\end{figure*}

We idealize the {\it Kepler} planets as being composed of ``solar
composition solids''; in other words, we assume they are bulk
chondritic and contain little H and He by mass. This assumption is
compatible with models of planetary interiors that reproduce the
observed radii and masses of close-in super-Earths.  Models comprising
gaseous H/He atmospheres overlying rocky cores are typically
characterized by low ($\lesssim 20$\%) gas fractions by mass (e.g.,
\citealt{rogersseager10}; \citealt{Lissaueretal11}; see also
\S\ref{sec_envelope}). Of course, merely knowing the radius and mass
of a planet is not sufficient to uniquely constrain its bulk
composition. In particular, worlds made predominantly of water are
also possible (e.g., \citealt{lopez12}). In this paper, we discount
water worlds on the grounds that water cannot condense {\it in situ}
in the hot inner regions of protoplanetary disks (but see \S\ref{sec_1214}
for a way to create close-in water worlds in a less strict
{\it in-situ} formation scenario).

Each of the {\it Kepler} planets is assigned a surface density 
\begin{equation}\label{eq_simple}
\sigma_{{\rm solid},i} \equiv \frac{M_i}{2\pi a_i \Delta a_i} \equiv \frac{M_i}{2\pi a_i^2}
\end{equation}
where $M_i = (R_i/R_{\oplus})^{2.06} \, M_\oplus$ (the best-fit
power-law mass-radius relation for the six Solar System planets
bounded in mass by Mars and Saturn;
\citealt{lissrag11}),
and $\Delta a_i = a_i = (P/{\rm yr})^{2/3} \, {\rm AU}$ is the semimajor axis
computed by assuming the host star has mass $M_\ast = M_\odot$.  
For
simplicity, we apply equation (\ref{eq_simple}) without regard to
whether a planet is solitary or is in a multi-planet system --- and have checked
that accounting for the relative spacings between planets when computing $\Delta a$
in multi-planet systems does not significantly change the overall distribution of inferred surface densities.

The $N = 1925$ surface densities $\{\sigma_{{\rm solid},i} \}$
so computed are displayed against
semimajor axes $\{a_i\}$ in Figure 2.  The median surface density
in each of 20 semimajor axis bins is shown as a red histogram. 
We fit a power law to the median data --- omitting the bin
at $a < 0.05$ AU because it contains systems that we consider
outliers (see \S\ref{sec:alternate} regarding our alternate MMEN construction; and also \S\ref{sec_extreme}).
This best-fit power law to the median data defines our standard
MMEN surface density in solids:
\be \label{eq_sigmad} \sigma_{\rm solid} = 6.2 \times 10^2 \,{\cal F}_{\rm disk}
\, \left( \frac{a}{0.2 \, {\rm AU}} \right)^{-1.6} \g \cm^{-2} \,,  
\ee
where ${\cal F}_{\rm disk} \geq 1$ accounts for how much more mass the
disk may have relative to the MMEN.  

Note that each $\sigma_{{\rm solid},i}$ does not depend on such
factors as ${\cal F}_{\rm transit}$ or ${\cal F}_{\rm detect}$ (see
\S\ref{sec:ubi}).  The set of $\{\sigma_{{\rm solid},i}\}$ merely
represents, by construction, the surface densities of the disks
required to form the known planets {\it in situ}.

Dividing $\sigma_{\rm solid}$
by $Z_{\rm solar} \times Z_{\rm rel}$, where $Z_{\rm solar} = 0.015$
is the total solar metallicity (Lodders 2003), we find a gas (H and
He) surface density of
\beq \label{eq_sigmag}
\sigma_{\rm gas} = 1.3 \times 10^5 \, {\cal F}_{\rm disk} \, \left( \frac{a}{0.2 \, {\rm AU}} \right)^{-1.6} \g \cm^{-2}\,.
\eeq

In the next two subsections, we consider alternate constructions
of the MMEN, using alternate input assumptions.

\subsection{Alternative construction of MMEN} \label{sec:alternate} A
disadvantage to using the median $\{\sigma_{{\rm solid},i}\}$ to
define the MMEN is that it does not account for relative occurrence
rates between semimajor axis bins. In particular, computing the median
in each bin does not reflect the fact that planets in the leftmost bin
at $a < 0.05$ AU are more rare than those in the neighboring bin (see
Figure 2) --- an observation which presumably implies real deficits of
disk surface density at $a < 0.05$ AU.  An alternate construction of
the MMEN that does account for such effects is made as follows. We
take the $N = 1925$ planets and sort them into semimajor axis bins
(assuming, as before, the period-semimajor axis relation for
solar-mass stars).  Each detected planet is augmented by $N_a = 200 \,
(a/{\rm AU})$ additional planets (of the same radius) to account for
the geometric probability of transit.  Planetary masses are estimated
according to the mass-radius relation cited above, and the surface
density in each semimajor axis bin is computed by adding together all
the mass in that bin and dividing by the annular area corresponding to
that bin.  The surface density profile is then normalized to a
per-star basis by dividing through by the total number of planets
summed over all bins (44307 = original plus augmented).

This alternative MMEN is shown as a black histogram in Figure 2.  For
the most part it tracks our standard MMEN, although it is everywhere
lower. At $a \gtrsim 0.05$ AU, we interpret the deficit to reflect
incompleteness in the Batalha et al.~(2012) catalog --- i.e., if
${\cal F}_{\rm detect}$ ranges from 60\% at $a \approx 0.2$ AU to 40\%
at $a\approx 0.4$ AU, then the alternate MMEN would come into
alignment with the standard MMEN. At $a \lesssim 0.05$ AU, it seems
unlikely that the Batalha et al.~(2012) catalog is incomplete (i.e.,
${\cal F}_{\rm detect} \approx 1$ at the smallest orbital distances),
and the factor of 10 difference between the alternate MMEN and the
standard MMEN probably underscores a real paucity of solid 
material there for most stars.  We propose that sublimation of dust is
responsible for this deficit (\S\ref{sec:astars}).

\subsection{Sensitivity of the normalization and slope of the MMEN surface
  density to input assumptions}\label{wulith12}

Taken at face value, the solid surface density $\sigma_{\rm
    solid}$ for our MMEN is a factor of 5 larger
  than the solid surface density of the MMSN --- see equation 2 of
  \citet{chiangyoudin10}, and use $Z_{\rm rel} = 0.33$ for the
  fraction of metals condensed as solids in the hot inner disk (i.e.,
  metals not taking the form of water; \citealt{lodders03}).  As a
  further point of comparison, our $\sigma_{\rm solid}$ is a factor of
  $7$ times larger than the surface density in rock
  as calculated by \citet{hayashi81}.

  These factors by which the MMEN is overdense compared to the MMSN
  may be mis-estimated as they are sensitive to our assumed
  mass-radius relation $M = M_\oplus (R/R_\oplus)^{2.06}$.  We
  experiment with two other mass-radius relations.  The first is from
  \citet{wulith12}, who deduced using transit timing variations that
  close-in super-Earths are better described by $M/M_{\oplus} \approx
  3 (R/R_\oplus)^1$ for $R \approx 1$--$7R_\oplus$. Using their
  mass-radius relation, we re-compute the MMEN and find $\sigma_{\rm
    solid} = 7.4 \times 10^2 \, (a / 0.2 \, {\rm AU})^{-1.8}$
  g/cm$^2$, which differs from our equation (\ref{eq_sigmad}) by about
  20\%. Looking at Figure 5 of \citet{wulith12}, we see that our
  fiducial $M \propto R^2$ relation still appears acceptable over the
  range $R = 2$--$5 R_\oplus$, which is where we have used it.

  Our second experiment uses the mass-radius relations calculated by
  \citet{rogersetal11} for rocky cores laden with hydrogen envelopes
  --- the kind of planets we expect from an {\it in-situ} formation
  scenario (\S\ref{sec_envelope}--\S\ref{sec_retain}).  Combining
  their Figure 4 with our expressions for envelope-to-core masses as
  derived in \S\ref{sec_envelope}, and repeating our calculation for
  the MMEN, we find that $\sigma_{\rm solid} \approx 1.7 \times 10^2
  \, (a / 0.2 \, {\rm AU})^{-1.6} \gm/{\rm cm}^{2}$ --- a result that
  practically coincides with the (extrapolated) MMSN.  The surface
  density drops by a factor of $\sim$4 relative to our standard model
  because the Rogers et al.~(2011) planets have extraordinarily low
  densities (they are ``puffy''); for example, a $3 R_{\oplus}$ planet
  with an H-to-rock fraction of 1\% might only have a mass of $1.5
  M_{\oplus}$ according to their Figure 4, because the hydrogen
  atmosphere is so voluminous.

Another way in which our calculation of $\sigma_{\rm solid}$
  is biased is in the incompleteness of the Batalha et al.~(2012)
  catalog, particularly for small planets. When we incorporate the
  incompleteness corrections calculated by Dong \& Zhu
  (\citeyear{dongzhu12}, see their Figure 5) for planets
  with $1 < R(R_\oplus) < 5$, thereby adding many
  small planets to our analysis, we find that $\sigma_{\rm solid} =
  4.3 \times 10^2 \, (a / 0.2 \, {\rm
    AU})^{-1.6} \gm/{\rm cm}^2$.

We conclude from these experiments that depending
  on the input assumptions, the factor by which
  the MMEN is overdense compared to the MMSN may range from $\sim$1 to
  $\sim$5, and that the power-law slope of the disk surface density
  with orbital radius ranges from -1.6 to -1.8 (compared to the
  canonical MMSN value of -1.5). From an order-of-magnitude
  perspective, the MMEN does not appear too different from the MMSN!
  Our fitted slope can be understood simply in terms of the
  observations.  Since $\sigma_{\rm solid} \propto M(a)/ (a \Delta a)$
  and $\Delta a \propto a$, it must be that the characteristic planet
  mass $M(a) \propto a^{0.3}$ if $\sigma_{\rm solid} \propto
  a^{-1.7}$. Indeed such an increase in planet mass with radius is
  evinced by the data in Figure 1 (and as already noted, the
  incompleteness corrections of Dong \& Zhu 2012 do not much affect
  this trend).

\subsection{Other physical properties of the MMEN} \label{sec:other}
At small stellocentric distances, the energy locally liberated by
accretion in a protoplanetary disk raises the midplane temperature
significantly above that of a disk passively heated by stellar
radiation. If the accretional energy is radiated vertically, the
factor by which the midplane temperature is boosted scales as
$\tau^{1/4}$, where $\tau$ is the Rosseland mean optical depth across
the vertical extent of the disk (e.g., \citealt{lecar06}, their equations 1 and 2).
Radiative equilibrium models of
actively accreting disks were constructed by, e.g.,
\citet{dalessioetal98} and \citet{dalessioetal01}, who found that as
the disk radius decreases, the midplane temperature tends to saturate
near the dust sublimation temperature of $\sim$1500 K; dust vaporizes
in hotter disks and throttles the optical depth back down.  The radial
extent of this near-isothermal zone depends sensitively on the assumed
grain size distribution (bigger grains yield smaller optical depths)
and on the disk surface density.  Note that the surface densities
(\ref{eq_sigmad})--(\ref{eq_sigmag}) of our MMEN are about two orders
of magnitude larger than those calculated by \citet{dalessioetal01}.

Given the uncertainties, we assume for simplicity an isothermal disk:
\begin{equation}
\label{eq_T}
T = 10^3 \K \,.
\end{equation}
For a disk of this temperature orbiting a solar mass star,\footnote{A disk
having a gas surface density of $\sigma_{\rm gas} = 1.3 \times 10^5$ g/cm$^2$
(equation \ref{eq_sigmag}) at $a = 0.2$ AU could have a midplane temperature of $T = 10^3$ K (equation \ref{eq_T}) if the mass accretion rate were $10^{-9} M_\odot$/yr and the disk opacity were $\sim$$0.03$ cm$^2$/(g of gas).
Such a disk opacity safely exceeds the lower bound
of $\sim$$10^{-3}$ cm$^2$/g which obtains when grains are completely
absent (\citealt{freedman08}).}
the hydrostatic thickness-to-radius aspect ratio is
\begin{equation}
\label{eq_h}
\frac{h_{\rm gas}}{a} = \frac{\cg}{\Omega a}= 0.03 \left( \frac{a}{0.2 \, {\rm AU}} \right)^{1/2} \\
\end{equation}
where $\cg$ is the gas sound speed and $\Omega$ is the Kepler orbital
frequency. Combining (\ref{eq_sigmag}) and (\ref{eq_h})
yields a midplane gas density
\begin{equation}
\label{eq_rho}
\rhog = \frac{1}{\sqrt{2\pi}} \frac{\sigma_{\rm gas}}{h_{\rm gas}} =
6 \times 10^{-7} \, {\cal F}_{\rm disk} \, \left( \frac{a}{0.2 \, {\rm AU}} \right)^{-3.1} \gm \cm^{-3} \,.
\end{equation}

The Toomre stability parameter \citep[e.g.,][]{binneytremaine} for
our gas disk is 
\begin{equation} \label{eq_Q}
Q_{\rm gas} = \frac{\cg \Omega}{\pi G \sigma_{\rm gas}} = 15 \, {\cal F}_{\rm disk}^{-1} \, \left( \frac{a}{0.2 \, \AU} \right)^{0.1} \,,
\end{equation}
large enough compared to unity that the MMEN is gravitationally stable.
Interestingly, if we allow ${\cal F}_{\rm disk} \approx 2$--3 (i.e.,
if planet formation were 30--50\% efficient), those systems
plotted in Figure 2 having the largest surface densities
$\{\sigma_{{\rm solid},i}\}$ have Toomre parameters $Q_{\rm gas}$
approaching unity (see also \S\ref{sec_extreme}). This suggests that
some planets formed in ``maximum-mass nebulae'' which were on the
verge of gravitational instability.

\section{ORDERS OF MAGNITUDE}\label{sec_oom}
We use the MMEN constructed in \S\ref{sec_mmen} to sketch how
close-in super-Earths can form {\it in situ}, working to order-of-magnitude accuracy.

\subsection{
Close-in super-Earths formed quickly, within gas disk lifetimes} 
\label{sec_fast}
Large disk surface densities $\sigma_{\rm solid}$ and short dynamical times
$\Omega^{-1}$ work together to enable close-in rocky planets to coagulate
rapidly. Discarding factors of order unity, we estimate that a planetary core of mass $M_{\rm core}$, radius $R_{\rm core}$, and bulk density $\rho_{\rm core}$ doubles its mass in a time
\begin{eqnarray}
t_{\rm coagulate}  =  \frac{M_{\rm core}}{\dot{M}_{\rm core}} & \sim & \frac{\rhocore R_{\rm core}^3}{\rhosolid \cdot {\cal F}_{\rm grav}R_{\rm core}^2 \cdot v_{\rm solid}} \\
 & \sim & \frac{\rhocore R_{\rm core}}{(\sigma_{\rm solid}/h_{\rm solid}) \,{\cal F}_{\rm grav}  v_{\rm solid}} \\
 & \sim & \frac{1}{{\cal F}_{\rm grav}} \frac{\rhocore R_{\rm core}}{\sigma_{\rm solid} \Omega}
\end{eqnarray}
where $\rho_{\rm solid}$ is
the mass density of the sea of planetesimals in which the planet
is immersed, $v_{\rm solid}$ is the velocity dispersion of planetesimals (assumed to be isotropic and greater than the planet's epicyclic velocity),
$h_{\rm solid} \sim v_{\rm solid}/\Omega$ is the vertical thickness of the planetesimal disk,
and ${\cal F}_{\rm grav} \geq 1$ is the factor by which gravitational focussing
enhances the accretion cross-section above its geometric value
(see \citealt{goldreichetal04} for a review). For our MMEN parameters,
planets form in short order (see also Montgomery \& Laughlin 2009 and \citealt{hansenmurray12}):
\begin{eqnarray}
t_{\rm coagulate} & \sim & \frac{2 \times 10^5}{{\cal F}_{\rm grav} } \left( \frac{\rhocore}{6 \gm \cm^{-3}} \right) \left( \frac{R_{\rm core}}{2R_{\oplus}} \right) \nonumber \\
& & \cdot \left( \frac{600 \gm \cm^{-2}}{\sigma_{\rm solid}} \right) \left( \frac{2\pi/\Omega}{30 \, {\rm days}} \right) \yr \label{eq_coag} \\
& \sim & \frac{2 \times 10^5}{{\cal F}_{\rm grav} } \left( \frac{R_{\rm core}}{2R_{\oplus}} \right) \left( \frac{a}{0.2 \,\AU} \right)^{3.1} \yr \,.
\end{eqnarray}
Since ${\cal F}_{\rm grav} \geq 1$, close-in super-Earths can readily form within the
lifetimes of inner gas disks,
estimated to be $t_{\rm gas} \sim 10^6$--$10^7 \yr$
from observations of near-infrared excesses of young stellar
objects (e.g.,
\citealt{hernandezetal08}).

We assume throughout this paper that close-in, rocky super-Earths form
to completion while the full, solar-abundance complement of gas is
still present. This assumption is not without peril.  Gas can
circularize the orbits of growing protoplanets by dynamical friction,
preventing orbits from crossing and halting further mergers. Under
these conditions, a protoplanet grows until it is ``isolated'' ---
i.e., when it has consumed all the solid material within $\sim$2.5
Hill radii of itself (\citealt{greenberg91}). We assume here that
dynamical friction cooling by gas does not prevent protoplanets from
merging beyond the isolation phase, and that coagulation completes
well before the gas disk dissipates --- indeed our predictions in
\S\ref{sec_envelope} for the amount of gas accreted by rocky cores
will rely on a full gas budget.  Our assumption could be justified if
the ambient disk is sufficiently turbulent that density fluctuations
in gas continuously excite protoplanet eccentricities (see, e.g.,
section 3.1 of the review by \citealt{kley12}), thereby maintaining a
chaotic sea of crossing orbits. The eccentricities required for
``isolation-mass'' embryos to cross orbits are on the order of
$\sim$0.01.

\subsection{
Close-in super-Earths accreted planetesimals each dozens of kilometers in size} \label{sec_planetesimal}

The planetesimals accreted by the protoplanets cannot be too small,
lest they spiral inward onto the star by aerodynamic drag. The
inspiral time, given by the planetesimal's orbital angular momentum
divided by the aerodynamic drag torque, must be longer than $t_{\rm
  coagulate}$: 
\beq \label{eq_inspiral} t_{\rm inspiral} \sim
\frac{\rhobulk s^3 \Omega a^2}{\rhog v_{\rm rel}^2 s^2 a} \, \gtrsim
\, t_{\rm coagulate} \eeq 
where $\rhobulk$ is the bulk density of a
planetesimal, $s$ is the planetesimal size, and $v_{\rm rel}$ is the
velocity of gas relative to the planetesimal ($v_{\rm rel}$ is assumed
anti-parallel to the planetesimal's orbital velocity).  In writing
(\ref{eq_inspiral}) we have assumed (and have checked {\it a
  posteriori}) that the drag force is appropriate for flow around a
blunt obstacle at high Reynolds number. To estimate $v_{\rm rel}$, we
recognize that gas is supported by a radial pressure gradient and
orbits the star at the sub-Keplerian velocity $(1-\eta)\Omega a$,
where $\eta \sim (h_{\rm gas} / a)^2 \sim 8 \times 10^{-4} (a/0.2
\,{\rm AU})^{1}$ (\citealt{adachietal76};
\citealt{chiangyoudin10}).  Taking the
planetesimal's orbital velocity to be the full Keplerian value, we
find that $v_{\rm rel} \sim \eta \Omega a \sim 50 \m \s^{-1} (a/
0.2\,{\rm AU})^{1/2}$. Thus to satisfy (\ref{eq_inspiral}), the
planetesimals must have sizes
\begin{eqnarray}
s & \gtrsim & \Omega t_{\rm coagulate} \frac{\rhog}{\rhobulk} \eta^2 a \\
  & \gtrsim & 50 \left( \frac{0.2\,\AU}{a} \right)^{1.6} \left( \frac{3 \g \cm^{-3}}{\rho_{\rm bulk}} \right) \left( \frac{t_{\rm coagulate}}{2 \times 10^5 / {\cal F}_{\rm grav} \, \yr} \right) \km  \, .
\end{eqnarray}

How are such super-km-sized planetesimals formed?  There is as yet no
consensus.  Recent work focuses on a combination of aerodynamic and
gravitational instabilities to agglomerate particles
(\citealt{youdingoodman05}; \citealt{johansenetal07};
\citealt{baistone10}; and references therein) --- or on pure gravitational
instability of a vertically thin and dense dust layer
\citep{leeetal10,leeetal10b,shichiang12}.

\subsection{
Close-in super-Earths accreted hydrogen from the primordial gas disk --- typically 3\% by mass, up to a maximum of $\sim$$1/(2Q_{\rm gas})$}
\label{sec_envelope}
Immersed in gaseous disks, rocky cores acquire gas
envelopes. A protoplanet derives an accretional luminosity from
infalling planetesimals, and the gas envelope must transport this
energy of accretion. Planetary accretion luminosities are so high (see
\S\ref{sec_fast}), and dust-laden gas in the dense inner disk
so optically thick, that the energy is likely
transported by convection --- i.e., the atmospheres are
adiabatic, not isothermal (e.g., \citealt{rafikov06,ikomahori12}). In
this case, as long as the gas envelope is not so extended that it
becomes truncated by stellar tides (see below), its mass scales with
the Bondi radius $\RB$:
\beq M_{\rm envelope} \sim 4\pi \rhog \RB^3
\eeq
where
\beq \RB = {GM_{\rm core}}/{\cg^2} \,.  \eeq
It follows that 
\beq \label{eq_gas:rock} \frac{M_{\rm envelope}}{M_{\rm core}}
\sim \left( \frac{M_{\rm core}}{M_{\rm core}^\ast} \right)^2 \sim 3\%
\left( \frac{M_{\rm core}}{4 M_{\oplus}} \right)^2 \left( \frac{23
    M_{\oplus}}{M^\ast_{\rm core}} \right)^2
\eeq
where
\beq
M^\ast_{\rm core} = \frac{\cg^3}{G^{3/2}(4\pi \rhog)^{1/2}} = 23 \, {\cal F}_{\rm disk}^{-1/2} \left( \frac{a}{0.2 \,\AU} \right)^{1.55} M_{\oplus} \,.
\eeq
For fixed $M_{\rm core}$, the gas-to-rock fraction $M_{\rm envelope}/M_{\rm core} \propto {\cal F}_{\rm disk} \, a^{-3.1}$.

The gas-to-rock fraction implied by equation (\ref{eq_gas:rock}) breaks down for large $M_{\rm core}$. Cores with 
\begin{equation} \label{eq_bondihill}
M_{\rm core} \gtrsim M_{\rm Bondi=Hill} = \frac{\cg^3
  a^{3/2}}{G^{3/2} (3M_\ast)^{1/2}} \approx 4 \left( \frac{a}{0.2\, \AU}
\right)^{3/2} M_{\oplus}
\end{equation}
have their gas envelopes truncated by stellar tides: for such massive cores, $\RB$ exceeds
\begin{equation}
\RH \sim \left( \frac{M_{\rm core}}{3M_\ast} \right)^{1/3} a
\end{equation}
the radius of the Hill sphere surrounding the planet beyond which stellar gravity circumscribes circumplanetary orbits. In this regime, the gas envelope
mass attains its maximum value set by $\RH$:
\begin{equation}
\max M_{\rm envelope} \sim 4\pi \rho_{\rm gas} \RH^3 
\end{equation}
so that
\begin{eqnarray} \label{eq_gas:rock1}
\max \frac{M_{\rm envelope}}{M_{\rm core}} & \sim & \frac{4\pi \rho_{\rm gas} a^3}{3M_{\ast}} \\
& \sim & 3\% \, {\cal F}_{\rm disk}^{1} \,\left( \frac{a}{0.2 \, \AU} \right)^{-0.1}
\end{eqnarray}
independent of $M_{\rm core}$. Having the planet's
  (adiabatically distended) gas envelope truncate at the Hill sphere
  --- which lies well inside the disk scale height $h_{\rm gas}$ for
  our parameters --- reassures that close-in super-Earths do
  not undergo runaway gas accretion to become gas giants.

Note that equation (\ref{eq_gas:rock1}) can be rewritten as
\begin{equation}
\max \frac{M_{\rm envelope}}{M_{\rm core}} \sim \frac{4}{3\sqrt{2\pi}} Q_{\rm gas}^{-1} \,. \label{eq_gas:rock1b} 
\end{equation}
Referring back to the curves of constant $Q_{\rm gas}$ overlaid
on Figure 2, we see that quite a few super-Earths (those gray points 
having the highest values of $\sigma_{{\rm solid},i}$)
could have gas-to-rock fractions of up to $\sim$30\%.

We caution that the equations in this subsection pertain to
adiabatic and not isothermal gas envelopes. Cores at larger disk radii
($a \gg 0.5$ AU) typically have lower accretional luminosities (lower
planetesimal accretion rates) and are immersed in less optically thick
gas. Radiative cooling efficiencies are consequently greater, and render
the disk at large $a$ --- e.g., at $a = 5$ AU where Jupiter resides
--- more conducive to runaway gas accretion.  See, e.g., Figure 7a of
\citet{rafikov06}. 

Clearly, the simple-minded analysis we have presented above for
gas-to-rock fractions does not capture a host of effects --- gas
cooling, spatial and temporal variations in the nebular pressure and
temperature beyond those we have assumed, and complicated
circumplanetary flows --- that could change our answers significantly
(see, e.g., \citealt{lubowdangelo06}; \citealt{ikomahori12}).  Our
intent is merely to show that {\it in-situ} accretion of gas envelopes
is {\it prima facie} plausible insofar as the order-of-magnitude
estimates of gas fractions in equations (\ref{eq_gas:rock}),
(\ref{eq_gas:rock1}), and (\ref{eq_gas:rock1b}) are in the same
ballpark as those inferred from radius-mass measurements of close-in
super-Earths (see, e.g., Figure 6 of \citealt{rogersetal11} --- but
note that these calculated present-day gas fractions are themselves
uncertain as they rely on opacities and atmospheric temperature
profiles that are not well constrained).

\subsection{
Many but not all close-in super-Earths
  retain their primordial hydrogen envelopes} \label{sec_retain}
The energy required for a hydrogen
atom of mass $m_{\rm H}$ to escape the atmosphere of a super-Earth is
$\sim$$GMm_{\rm H}/R \sim 2 \,(M/10 M_\oplus) \,(3
R_{\oplus}/R) \,\eV$, large enough that only high-energy X-ray and UV (XUV)
photons from the parent star
can impart the requisite energy by photoionization (by comparison,
stellar optical radiation heats particles up to a mean energy $kT_{\rm eff} \sim 0.06 \, \eV$, where $k$ is Boltzmann's constant
 and $T_{\rm eff} \sim 700\,\K$ is the effective blackbody temperature of the planet).
We assume that a fraction $\epsilon$ of the impinging XUV stellar
radiation goes toward lifting gas out of the planet's gravity
well:
\begin{equation} \label{eq_energy_lim}
\epsilon \frac{L_{\rm XUV}}{4\pi a^2} \cdot \pi R^2 = \frac{GM\dot{M}}{R} \,.
\end{equation}
Mass loss in this regime is ``energy-limited'' (e.g., \citealt{murrayclay09}).
We adopt
\begin{equation}
L_{\rm XUV} \approx 3 \times 10^{-6} \left( \frac{t}{5 \, {\rm Gyr}} \right)^{-1.23} L_{\odot}
\end{equation}
for $t > 0.1$ Gyr \citep{ribas05}, and
\begin{equation}
L_{\rm XUV} \approx 3 \times 10^{-4} L_{\odot}
\end{equation}
for $t < 0.1$ Gyr. We further take $R = 5 R_\oplus$ at $t < 0.1$ Gyr
to account for how the primordial gas envelope may be distended because of an initially high entropy (cf.~\citealt{lopez12}), and set $R = 3 R_\oplus$ for $t > 0.1$ Gyr.
Given these assumptions, most of the mass loss occurs at $t < 0.1$ Gyr,
and the amount of mass lost is
\begin{equation} \label{eq_masslost}
\Delta M_{\rm envelope} \sim 0.01 \left( \frac{\epsilon}{0.1} \right) \left( \frac{R}{5 R_\oplus} \right)^3 \left( \frac{10 M_\oplus}{M} \right) \left( \frac{0.2 \, \AU}{a} \right)^2 M_{\oplus} \,.
\end{equation} 
This is a factor of 10--100 less than the hydrogen envelope masses
imputed to many (but not all) {\it Kepler} super-Earths based on
radius-mass measurements, and it is similarly smaller than the
envelope masses we estimated in \S\ref{sec_envelope}.  Thus we expect
many super-Earths to largely retain their hydrogen envelopes. There
is, however, enough variance in the input parameters that we would
also expect some planets (having some combination of small $a$, small
$M_{\rm envelope}$, or large incident XUV flux) to be stripped clean
of their hydrogen veneers.\footnote{Giant impacts may also erode
  atmospheres; see \S\ref{sec_stay}.}  Individual cases are discussed in
\S\ref{sec_extreme}.

\subsection{
Close-in super-Earths stay in place after formation}
\label{sec_stay}
Super-Earths formed at small stellocentric distances have not much choice
but to stay there. They cannot scatter each other out of the
gravity well of the star, because the largest velocity dispersion
they can attain by mutual scatterings is set by their mutual surface
escape velocity
\begin{equation}
v_{\rm esc} \sim 20 \left( \frac{M_1 + M_2}{20 M_\oplus} \right)^{1/2} \left( \frac{ 6 R_\oplus }{ R_1+R_2} \right)^{1/2} \km \s^{-1}
\end{equation}
which is less than the escape velocity from the star
\begin{equation}
v_{{\rm esc},\ast} \sim 90 \left( \frac{0.2 \, \AU}{a} \right)^{1/2} \km \s^{-1} \,.
\end{equation}
Thus dynamical instabilities --- i.e., planet-planet scatterings ---
in the inner disk do not result in ejection but rather in
planet-planet mergers (or, in rare instances, tidal interactions with
the star and orbital decay).

{\it In-situ} formation predicts that an ``oligarchy'' composed of
closely nested protoplanets eventually destabilizes when the oligarchs
viscously stir themselves faster than the underlying disk can cool
them by dynamical friction \citep[e.g.,][]{kokuboida00,
  goldreichetal04, kenyonbromley06, fordchiang07}. Oligarchs
ultimately scatter each other onto crossing orbits. The inequality
$v_{\rm esc} < v_{{\rm esc},\ast}$ implies that in the ensuing chaos,
protoplanets coalesce --- with the mass-doubling time $t_{\rm coag}$
given by the relations in \S\ref{sec_fast}. The total number of bodies
decreases and the rate of scattering declines. The final set of
super-Earths will occupy, at first, 
orbits with eccentricities
and inclinations up to $v_{\rm esc}/v_{{\rm esc},\ast} \sim 0.2$;
these will be circularized and flattened by any residual disk of
planetesimals (\S\ref{sec_circ}). The leftover disk of solids will, in
turn, eventually be consumed by the planets, at least in part.

Since ambient gas does not accrete in runaway fashion onto protoplanet
cores (\S\ref{sec_envelope}), and since photoevaporation is not
efficient at removing gas at small orbital radii (gas cannot be heated
to the temperatures of $\sim$$10^6 \K$ required for $\cg$ to approach
$v_{{\rm esc},\ast}$), gas that is not bound to cores in the inner
disk has no option but to accrete onto the star. The mechanism of disk
accretion is uncertain, but the magnetorotational instability is
arguably viable in these hot and relatively well-ionized regions
(cf.~\citealt{bai11} and \citealt{perezbecker11b,perezbecker11a} for
difficulties with appealing to the magnetorotational instability for
the rest of the disk).

\subsection{
Close-in super-Earths currently reside in co-planar and nearly circular orbits}
\label{sec_circ}

Once disk gas dissipates to the point where its turbulent density
fluctuations no longer perturb super-Earths significantly
(\S\ref{sec_fast}), the remaining planets have their orbital
inclinations and eccentricities damped by dynamical friction with
leftover planetesimals. The velocity dispersion of residual
planetesimals is, in turn, damped by inelastic collisions and/or drag
by remaining gas. The leftover disk of solids need only be a small
fraction of the original disk of solids to circularize and flatten
planetary orbits. An initial planetary eccentricity $e_{\rm init}$
damps to zero over a time
\begin{eqnarray}
t_{\rm circularize} & \sim & 0.1 \,\frac{M_\ast}{M} \frac{\Omega a}{G \sigma_{\rm leftover}} e_{\rm init}^4 \\
& \sim & 2 \times 10^5 \left( \frac{10\,M_\oplus}{M} \right) \left( \frac{6 \gm \cm^{-2}}{\sigma_{\rm leftover}} \right) \nonumber \\
& & \cdot \left( \frac{a}{0.2\,\AU} \right) \left( \frac{e_{\rm init}}{0.1} \right)^4 \yr 
\end{eqnarray}
(e.g., \citealt{fordchiang07}; the pre-factor of 0.1 contains a
Coulomb-type logarithm). An analogous formula applies for damping the
mutual inclination between the planetary orbit and the disk of planetesimals. 

Because eccentricities and inclinations are so readily damped by
leftover planetesimals, we expect close-in super-Earths to
occupy nearly circular, coplanar orbits.  A hard lower bound on the
final eccentricities will be established by gravitational interactions
between the planets {\it in vacuo}.
A lower bound on the mutual inclinations is less clear, as a system
that is perfectly coplanar will remain so forever.  The final
eccentricities and inclinations will depend not only on how strongly
planets stir one another during the post-oligarchic phase of planet
formation --- and the stirring rate is a sensitive function of
oligarchic mass and spacing (\citealt{chambers96}; Ford \& Chiang
2007) --- but also on the rate at which the leftover disk dissipates.
At least in the case of the Solar System's giant planet satellites,
eccentriticies are typically of the same order of magnitude as
inclinations.


\subsection{Some extreme extrasolar systems} \label{sec_extreme} We
consider now some specific examples of super-Earths that have garnered
special attention in recent years.  We apply the order-of-magnitude
analyses in previous subsections to assess the possibility that these
planets formed {\it in situ}. Some properties of these planets  
will be found to lie at the extremities of parameter space,
  and we will identify arguments both for and against
  {\it in-situ} accretion.

\subsubsection{The Kepler-11 system (b--f)} \label{sec_11}

The innermost five planets of the Kepler-11 system have a total mass
of $M_{\rm tot} \approx 35 M_\oplus$ between $a_1 = 0.09$ AU and $a_2
= 0.25$ AU around a solar-mass star \citep{Lissaueretal11}. If we
assume that the planets are composed predominantly of rock (see Figure
5 of \citealt{Lissaueretal11}), then the solid surface density of the
``minimum-mass Kepler-11'' system is $\sigma_{\rm solid} \sim M_{\rm
  tot} / 2\pi a_2^2 \sim 2.4 \times 10^3 \gm \cm^{-2}$, about 6 times
larger than that of the MMEN at $a = 0.25$ AU. For a solar metallicity
disk, the Toomre parameter $Q_{\rm gas} \sim 2.5$, suggesting that the
Kepler-11 primordial disk may have been on the verge of gravitational
instability --- i.e., it may have been a ``maximum-mass'' nebula.  The
packed set of planets is reminiscent of an oligarchy (e.g.,
\citealt{kokuboida00}; \citealt{goldreichetal04}).

Substituting the parameters of the Kepler-11 disk into equations (\ref{eq_gas:rock})--(\ref{eq_gas:rock1b}), we expect the primordial
H-He mass fraction of a Kepler-11 planet to range from
\begin{equation}
\frac{M_{\rm envelope}}{M_{\rm core}} \sim 4\% \left( \frac{M_{\rm core}}{2 M_{\oplus}} \right)^2 
\left( \frac{a}{0.2 \, {\rm AU}} \right)^{-3.1}
\end{equation}
for $M_{\rm core} < M_{\rm Bondi=Hill} \approx 4 (a/0.2 \, {\rm AU})^{3/2} M_\oplus$ (this is the case for planet f),
to a maximum gas fraction of
\begin{equation} \label{eq_kepler11}
\max \frac{M_{\rm envelope}}{M_{\rm core}} \sim 20\% \left( \frac{2.5}{Q_{\rm gas}} \right)
\end{equation}
for $M_{\rm core} > M_{\rm Bondi=Hill}$ (planets b, c, d, and
  e).  Our estimates of primordial gas-to-rock fractions are
consistent to order-of-magnitude with interior models that assume gas
envelopes overlying rocky cores: to wit, the interior models for
Kepler-11c, d, e, and f are characterized by gas-to-rock fractions of
2--20\% \citep{lopez12}.  

An outlier is Kepler-11b, which has the smallest inferred H
  fraction of $M_{\rm envelope}/M_{\rm core} \sim 0.05$--1.4\%
  (\citealt{lopez12}). At least in principle, this small amount of H
  may be what remains of a once-massive atmosphere 
  whittled down by XUV photoevaporation. From equation
  (\ref{eq_kepler11}) we expect that Kepler-11b could have once had an
  atmosphere amounting to $\sim$20\% of the total planet mass, while
  equation (\ref{eq_masslost}) predicts that XUV radiation could have
  eroded away $\Delta M_{\rm envelope} > 0.1 M_\oplus$ or $>$2\% of
  the planet mass --- this is a lower limit because Kepler-11b could
  have undergone ``runaway'' mass loss during which the planet stayed
  inflated for long duration \citep{baraffe04}. In fact, Lopez et
  al.~(2012) surmised that if Kepler-11b formed stricly {\it in situ}
and is not a water world, then it started with an enormous initial
gas fraction of $\sim$90\% in order to retain its currently thin
  veneer of hydrogen; these authors actually argued against {\it
    in-situ} formation because such a large initial planet mass would
  have threatened the orbital stability of the system.  More work
  is required before the case of Kepler-11b is resolved; the large
  parameter space of atmospheric opacities, cooling histories, and
  planetary magnetic fields (\citealt{adams11}) would seem to allow a
  wide range of mass loss histories.\footnote{The possibility
    that Kepler-11b's present-day atmosphere results from ongoing
    outgassing is not viable, because outgassing requires chemical
    reactions between surface iron and water, neither of which is
    available. Water is not available because water should not
    condense in the hot inner nebula, and iron is not available at the
    planet surface because it is rapidly sequestered in the planetary
    core soon after the planet's formation. These same remarks apply
    for GJ 1214b (\S\ref{sec_1214}).}  


\subsubsection{Kepler-10b} \label{sec_10}

Kepler-10b is a rocky super-Earth of mass $M = 4.56 M_\oplus$, $R =
1.42 R_\oplus$, and bulk density $\rho = 8.8^{+2.1}_{-2.9} \gm
\cm^{-3}$, orbiting a solar-mass star with a period of 0.836 days and
a semimajor axis $a = 0.0168$ AU \citep{batalha11}. Our estimate
of the surface density of the ``minimum-mass Kepler-10b'' disk is
$\sigma_{\rm solid} \sim M / 2 \pi a^2 = 7 \times 10^4 \gm
\cm^{-2}$.  This is 2--3 orders of magnitude larger than the surface
density of the MMEN in the smallest semimajor axis bin ($a < 0.05$ AU;
Figure \ref{fig2}) and is a measure of just how unusual Kepler-10b is.
Nevertheless, the Toomre parameter for the corresponding solar
metallicity disk is $Q_{\rm gas} \sim 6$, large enough for the
primordial gas disk to remain gravitationally stable.

Kepler-10b's large bulk density --- between that of
  compressed iron and a bulk Earth-like composition \citep{batalha11}
  --- indicates that it does not have as much C and O as would be
  expected for a planet that congealed outside the methane/water
  ice-lines in the primordial disk. Thus the inferred bulk composition
  of Kepler-10b argues against wholesale orbital migration.

The mass-doubling time for Kepler-10b is astonishingly
short (see equation \ref{eq_coag}):
\begin{equation}
t_{\rm coagulate} \sim 30 \, (R/1.4 R_\oplus) \, {\cal F}^{-1}_{\rm grav} \yr \,.
\end{equation}
Most of the time building Kepler-10b was probably spent forming the
seed planetesimals (via an unknown mechanism;
\S\ref{sec_planetesimal}).  Indeed $t_{\rm coagulate}$ is so short
compared to the typical lifetimes of gas disks ($t_{\rm gas} \sim
10^6$ yr) that the planet almost certainly did not assemble
early in the gas disk's evolution; so much angular momentum and mass
must have been transported across the inner disk that it is hard to
imagine how early-formed planets could have avoided being torqued by
the disk. An {\it in-situ} accretion scenario for fast-growing objects
like Kepler-10b appears to demand that they form at the very end of
the gas disk's life.

Kepler-10b is situated so close to its parent star that stellar XUV
radiation likely removed any hydrogen atmosphere it accreted from the
parent disk. Plugging the parameters of Kepler-10b into equation
(\ref{eq_masslost}), we see that enough XUV radiation was deposited to
remove up to $\Delta M_{\rm envelope} \sim 2 M_\oplus$ or $\sim$40\%
of its total mass. Kepler-10b probably never had this much hydrogen.
The planet's mass exceeds $M_{\rm Bondi=Hill} = 0.1
(a/0.017\,\AU)^{3/2} M_\oplus$ (equation \ref{eq_bondihill}), and
therefore its primordial gas fraction was at most $\sim$9\%
$(6/Q_{\rm gas})$ (equation \ref{eq_gas:rock1b}). Thus it is not
surprising that Kepler-10b can be modeled today as a pure rock planet.

\subsubsection{GJ 1214b} \label{sec_1214}

GJ 1214b is a super-Earth of mass $M = 6.55 M_\oplus$, $R = 2.68
R_\oplus$, and bulk density $\rho = 1.87 \pm 0.4 \gm \cm^{-3}$,
orbiting an M dwarf of mass $M_\ast = 0.157 M_\odot$ and luminosity $L_\ast = 3 \times 10^{-3} L_\odot$ with a period of
1.58 days and a semimajor axis $a = 0.014$ AU
\citep{charbonneau09}.  An estimate of the surface density of the
``minimum-mass GJ 1214b'' disk is $\sigma_{\rm solid} \sim M / 2\pi
a^2 = 1.4 \times 10^5 \gm \cm^{-2}$. This surface density is not readily
compared against the MMEN because the MMEN is constructed as an average
over solar-mass stars, not M dwarfs like GJ 1214.  The Toomre parameter 
for the corresponding solar metallicity disk is
$Q_{\rm gas} \sim 1.4$ --- possibly signaling a system that spent some time
on the brink of gravitational instability.

Differences in the mean molecular weight of an atmosphere
  lead to differences in atmospheric scale height, and these can be
  distinguished by measuring transit depths as a function of
  wavelength \citep{millerricci09}.  Such measurements have been made
  of GJ 1214b, from wavelengths $\lambda =
  0.6\,\mu$m to $5\,\mu$m \citep{Beanetal10,Desertetal11,Crolletal11,
    Beanetal11, Bertaetal12, deMooijetal12}. Most of these
  observations indicate no significant variation of transit depth with
  wavelength, which {\it prima facie} points to a massive and high
  molecular weight atmosphere (made, e.g., of CO$_2$; Lloyd \&
  Pierrehumbert 2013, submitted). This is seemingly incompatible with strict
  {\it in-situ} accretion.
  But in a less strict {\it in-situ} scenario, sizable ice-rich
  planetesimals could have migrated inward from outside the
  carbon/oxygen condensation radii in the primordial disk and
  assembled into close-in planets at their current orbital distances
  (Hansen \& Murray 2012). Planetesimals that are large enough (i.e.,
  much larger than dust grains) would have sublimation times exceeding
  coagulation times and would thus be safe against vaporization.

Yet another interpretation is that the atmosphere actually does
have a significant H component, but that its flat transmission spectrum 
arises from hazes (made possibly of
  KCl or ZnS; Morley et al., in preparation). More data are forthcoming for
  GJ 1214b (there is, e.g., a 60-orbit {\it Hubble Space Telescope} campaign led
  by J.~Bean), and whether some form of {\it in-situ} accretion
  can be reconciled with the data remains to be seen.


\section{PROSPECTS AND PREDICTIONS}\label{sec:conc}
Over the past two decades, extrasolar planets have 
completed their migration from the fringe to the mainstream of
astronomy. Time and again, they have defied theoretical
predictions regarding their properties in nearly every corner of
parameter space. It may be that the community's state of near-continual surprise
stems from an ingrained appeal to our Solar System as our standard template.
In this heliocentric view, the default position is to regard
extrasolar systems as exotica. We have shown, however,
that the close-in super-Earths detected in abundance
by the {\it Kepler} mission and Doppler velocity surveys
cannot be viewed this way: more than half, if not nearly all Sun-like stars
in the Galaxy harbor planets with radii $2 R_\oplus < R < 5 R_\oplus$
and periods $P < 100$ d. Super-Earths are not anomalous; they are the rule
that our Solar System breaks. In a sense,
the burden of explaining planetary system architectures
rests more heavily on the Solar System than on the rest of the Galaxy's
planet population at large.

The omnipresent close-in super-Earths enable us to construct a new
template, the minimum-mass extrasolar nebula (MMEN), which in turn can
be used to explore the possibility that such planets formed {\it in
  situ}. Our order-of-magnitude sketches in this regard are
promising. {\it In-situ} formation at small stellocentric distances
has all the advantages that {\it in-situ} formation at large
stellocentric distances does not: large surface densities, short
dynamical times, and the deep gravity well of a parent star that keeps
its planetary progeny in place.\footnote{Another way to say this is
  that ``particle-in-a-box'' simulations of planetary coagulation are
  adequate in the innermost regions of protoplanetary disks, as
  material is confined to the ``box'' of the star's potential well.}
The large surface densities of the MMEN, if extrapolated to
  greater stellocentric distances, would also help to form distant gas
  and ice giants --- for a theoretical perspective, see the disk-mass
  enhancement factors invoked to form Jupiter (e.g.,
  \citealt{lissaueretal09}) and Uranus and Neptune (e.g.,
  \citealt{goldreichetal04}), and for an observational view, see the
  microlensing surveys which report order-unity occurrence rates for
  super-Earths and gas giants from 0.5--10 AU (\citealt{gaudi12}, his
  section 6.2).  The biggest challenge for theories of planet
formation --- and this is true regardless of whether planets migrated
or not --- is in understanding how seed planetesimals form, i.e., how
objects grow from sub-cm sizes to super-km scales.

Our unfinished theories of planetesimal formation notwithstanding, the
basic properties of close-in super-Earths that form at their current
orbital distances seem clear. {\it In-situ} formation with no
large-scale migration generates short-period planets with a lot of
rock and metal and very little water. The accretion of nebular gas
onto protoplanetary cores of metal produces H/He-rich atmospheres of
possibly subsolar metallicity that expand planets to their observed
radii. Retainment of primordial gas envelopes against photoevaporation
leads to planets that can be similar in bulk density to Uranus and
Neptune while being markedly different in composition.  Close-in
planets are not water worlds (but see the discussion
under \S\ref{sec_1214} for a way to produce close-in water worlds
in a less strict {\it in-situ} formation scenario).

We are all too aware that predictions in the subject of planet
formation have a poor track record. Nevertheless, in the firm
conviction that good theories are falsifiable ones, we offer here a
set of observationally testable consequences of {\it in-situ}
formation, with the aim of bringing the ongoing debate about planetary
origins into sharper focus. As we will see, the prospects for further probing
many of our ideas are bright.

\subsection{Early type stars should lack close-in rocky super-Earths}\label{sec:astars}
The {\it in-situ} formation hypothesis depends on the
availability of dust to form seed planetesimals.
Where it is too hot for dust to survive, there should be no
planets.

How does the ``dust-line'' (i.e., the radius inside of which dust is
vaporized) vary with stellar mass?  Near-infrared interferometric
measurements indicate that dust-lines expand with increasing stellar
luminosity, from 0.05--0.1 AU in the case of T Tauri stars, to
0.1--0.5 AU for Herbig Ae stars, to 0.5--10 AU for Herbig Be stars
(see Figure 7 of the review by \citealt{DullMonn10}). Under {\it in-situ}
formation, the prediction is clear: hotter stars should have fewer
planets closer in.

Dust sublimation offers a simple explanation of the dramatic
drop in the occurrence rate of super-Earths at $a < 0.05$ AU (\S\ref{sec:alternate}). The drop at $a < 0.05$ AU applies to Sun-like
stars, whose progenitors are closest to the T Tauri population.

Our prediction that main-sequence A and B stars --- the descendants of
Herbig Ae and Be stars --- host fewer close-in super-Earths finds some
preliminary support in the {\it Kepler} data.  Among stars in the {\it
  Kepler} Input Catalog for which (1) photometric light curve data
exist, (2) estimates of stellar effective temperature $T_{\rm eff}$
are tabulated, and which are (3) not flagged as red giants, we find
that only 2 out of 1965 stars with $7500\,{\rm K}<T_{\rm
  eff}<10000\,{\rm K}$ harbor planetary candidates with
$2.5\,R_{\oplus} < R < 5\,R_{\oplus}$. On the other hand, among
131,987 primaries with $T_{\rm eff}<6000\,{\rm K}$, there are 549
candidate planets with $2.5\,R_{\oplus} < R < 5\,R_{\oplus}$ --- a
rate that is more than four times higher than the planetary occurrence rate
for presumably early-type stars. While this comparison is
crude and ignores a host of selection and other potential biases, it
is nevertheless consistent with our expectation, and motivates an
improved analysis.

\subsection{Brown dwarfs and M dwarfs should be accompanied by close-in super-Earths and Earths}

If the resemblance noted in \S\ref{sec:sat} between close-in
super-Earths and giant planet satellites is not a coincidence, then
simple interpolation leads to the expectation that stars at the bottom
of the main sequence, and brown dwarfs, will be commonly accompanied
by planets (satellites) with masses $M \sim 10^{-5}$--$10^{-4}
M_{\ast}$ and $P \lesssim 100$ days. Such low-luminosity primaries are
outstanding hosts for {\it in-situ} planet formation for the same reason
that massive early-type stars are not --- dust can only survive, and
by extension planets can only exist, where disks are cool enough.

For a red dwarf primary of mass $M_\ast \sim 0.1 M_{\odot}$ and
$R_\ast \sim 0.1 R_\odot$, we might expect planet masses $M \sim
10^{-5}$--$10^{-4} M_\ast \sim 0.3$--$3 M_{\oplus}$ and 
radii $R \sim 0.7$--$1.5 R_\oplus$. The corresponding transit depths are
encouragingly large, $(R/R_{\ast})^2 \sim 0.004$--$0.02$.  The MEarth
Project \citep{NutzChar08} is an ongoing transit survey of up to
$\sim$2000 M dwarfs.
The geometric probability of transit for
a $P = 10\,{\rm d}$ companion orbiting a 0.1$M_{\odot}$ primary is
${\cal F}_{\rm transit} \sim 1\%$, suggesting that up to a few dozen transiting
systems might eventually be detected.
The MEarth Project currently sets exposure times to detect planets
with $R = 2R_{\oplus}$ at 3$\sigma$ confidence
\citep{Bertaetal12}.  If the naive and simple-minded scalings we
have discussed above hold, then exposure times
would need to be adjusted upward to gain access to the bulk of the planets
arising from {\it in-situ} formation and having $R < 2 R_\oplus$.

\subsection{Stellar binaries can be used to rule out migration}
Planets cannot form at large distance from a star if that star is
orbited by a close enough companion. Gravitational perturbations from
a stellar companion can destabilize certain swaths of circumstellar
space. In destabilized regions, planetesimals can be excited to such
large velocity dispersions that they erode upon colliding.  For
example, in the protoplanetary disk of Alpha Centauri B, perturbations
from star A on its $P = 80$ yr, $e = 0.5$ orbit can thwart planet
formation by grinding planetesimals to dust at disk radii $a \gtrsim
0.5$ AU (\citealt{Thebaultetal09}, cf.~\citealt{rafikov12} who notes
that the boundary for erosive collisions depends also on disk self-gravity). Were planets ever to be discovered orbiting Alpha Cen B
just inside the $a \approx 0.5$ AU boundary, they would be
incompatible with long-distance inward migration.  The prospects for
using stellar binaries to distinguish between {\it in-situ} formation
and migration appear encouraging; \citet{dumusque12} have just
announced the discovery of a 1.1 $M_\oplus$ planet in an orbit with
semimajor axis $a \approx 0.04$ AU about Alpha Cen B.

Objects at the bottom of the main sequence that are members of wide
stellar binaries can undergo {\it one-time} eclipses during the course
of large-scale photometric surveys.  Should the orbital geometry
cooperate, a low-mass star can haul its retinue of satellites across
the face of a more luminous star.  The slow pace of the eclipse (e.g.,
a central transit of a $1 M_{\odot}$ primary by a companion with $a=5$
AU lasts 30 hours) improves detection signal-to-noise for satellites
of the low-mass secondary. A back-of-the-envelope estimate for the
number of one-time eclipses of G dwarfs by low-mass (M dwarf or brown
dwarf) companions that {\it Kepler} might observe is
\begin{equation}
N_{\rm one-time} \approx 9 
\left( \frac{{\cal F}_{\rm binary}}{0.6} \right)
\left( \frac{{\cal F}_{\rm period}}{0.1} \right) 
\left( \frac{{\cal F}_{\rm transit}}{10^{-3}} \right) 
\left( \frac{N_{\ast,{\rm total}}}{1.56 \times 10^5} \right) \,,
\end{equation}
where ${\cal F}_{\rm binary}$ is the fraction of G dwarfs with
low-mass companions; ${\cal F}_{\rm period}$ is the fraction of such
binaries with orbital periods between $\sim$3 and 30 yr; and ${\cal
  F}_{\rm transit}$ is the transit probability normalized to a binary
separation of 5 AU $\sim 10^3 R_\ast$ \citep{DuqMayor91}.  During each
of these one-time eclipse events, transits of the G dwarf by planets
orbiting the low-mass dwarf are guaranteed if the planets' orbital
planes are nearly aligned with the stellar binary orbital plane.  Even
if a given planetary plane is randomly oriented with respect to the stellar binary plane, the probability of
catching a planetary transit drops only by a factor of $\sim$$\pi/9$.
For the proposed {\it Transiting Exoplanet Survey Satellite} (TESS) Mission \citep{Rickeretal10}, which will survey $N_{\ast,{\rm total}}
\sim 2 \times 10^6$ stars, $N_{\rm one-time} \sim 10^2$.

\subsection{Orbit normals of close-in super-Earths should be aligned with stellar spin axes}
Basic considerations of angular momentum dictate that close-in
protoplanetary disks, and the close-in planets that they breed,
should have orbital planes that are well-aligned with the equator planes of
their host stars.\footnote{ Of course, smooth disk-driven migration can also produce spin-aligned orbits. But smooth convergent migration is also expected to produce a preponderance of planets locked in mean-motion resonances,
and this is not observed (for a variety of perspectives on this issue, see \citealt{LithWu12}; \citealt{BatMor12}; \citealt{rein12}; \citealt{petrovich12}).} This is an eminently testable prediction for {\it Kepler} planet-bearing systems. Although
Rossiter-McLaughlin measurements are too difficult to make for most
super-Earths, the stellar equatorial inclinations $i_{\ast}$ relative to the sky plane can be determined by conventional means \citep{Hiranoetal12}. A single high-resolution, high
signal-to-noise spectrum can be used to obtain a 
given star's projected rotational velocity
$v_{\rm rot} \sin i_{\ast}$ and radius $R_{\ast}$. In addition, the {\it Kepler}
light curve yields the stellar rotation period
$P_{\rm rot}$. With such measurements in hand,
\begin{equation}
  i_{\ast}=\arcsin\left(\frac{P_{\rm rot} \cdot v_{\rm rot} \sin i_\ast}{2\pi R_\ast}\right) \,.
\end{equation}
{\it In-situ} formation predicts that transiting planets should orbit stars
for which $i_{\ast}\sim 90^{\circ}$. Because only one spectrum is required
per star, this test can be carried out for a large number of targets.

\subsection{Close-in super-Earths have primordial H/He atmospheres}
Close-in ``water worlds'' --- planets with substantial components of
O, C, and N-rich ices, like Neptune\footnote{Many papers
    distinguish between ``water worlds'' with water-rich surface
    layers and ``hot Neptunes'' with hydrogen-rich
    atmospheres. Throughout our paper we do not follow this practice
    because Neptune actually contains a thick mantle of water,
    ammonia, and methane ices, and thus could arguably be called a
    ``water world''. The distinction we make in our paper is between
    ``water worlds'' (which are CNO-rich in bulk composition,
    including Neptune) and the more general ``super-Earths''.} ---
are ruled out under strict {\it in-situ} formation because such ices
could not have condensed in the hot inner regions of protoplanetary
disks (but see \S\ref{sec_1214} for a less strict scenario).
We have shown that close-in super-Earths embedded in their parent gas
disks can accrete H/He-rich gas envelopes weighing from a few percent
to tens of percent of the planet mass (\S\ref{sec_envelope}). We have
also argued that many but not all such planets retain their
atmospheres against photoevaporation (\S\ref{sec_retain}). 
  These expectations are currently in tension with Kepler-11b
  (\S\ref{sec_11}) and GJ 1214b (\S\ref{sec_1214}).  Spectroscopic
  measurements of GJ 1214b, HD 97658b, and 55 Cnc e are forthcoming
  with the {\it Hubble Space Telescope} (Cycle 20, PI: J.~Bean).
More super-Earths should also be discovered transiting bright
primaries and be amenable to spectroscopic follow-up.  For example,
\citet{Bonfilsetal12} have recently announced the detection of GJ
3470b, a planet with $M =14M_{\oplus}$ and $R=4.2 \,R_{\oplus}$
orbiting a spectral type M1.5 V host.


\subsection{Super-Earths that retain primordial gas atmospheres should be centrally concentrated with small $k_2 \lesssim 0.05$}
A rocky planet with an extended H atmosphere is more centrally
condensed than a water world with a volatile-rich mantle or
atmosphere.  Higher degrees of central condensation map to smaller
values of the tidal Love number, $k_2$. To get a sense of what
$k_2$-values we may expect for close-in super-Earths, we look to the
interior models by \citet{kramm11} of GJ 436b ($M \approx 22
M_\oplus$, $R \approx 4.3 M_\oplus$, $a \approx 0.03$ AU).  For pure H
atmospheres overlying massive cores with $M_{\rm core} / M > 80$\% ---
these H-atop-rock solutions are the ones favored by our {\it in-situ}
formation scenario --- Kramm et al.~(2011) calculated that $k_2 <
0.05$. By contrast, for Neptune (which consists of an H atmosphere
draped over a thick water layer, which in turn overlays a rocky core
for which $M_{\rm core}/M < 0.25$), $k_2 \approx 0.16$. Our
expectation that $k_2 \lesssim 0.05$ applies only to those close-in
super-Earths that retain the sizable H envelopes
(\S\ref{sec_envelope}) accreted from the primordial gas
disk. Those few pure-rock close-in super-Earths whose atmospheres were
obliterated by photoevaporation have relatively homogeneous interiors
and thus larger $k_2$ (e.g., for the Earth, $k_2 \approx 0.3$).

In a tidally evolved, two-planet system, it is possible to infer
$k_2$ for the inner planet \citep{Batyginetal09}.  Tidal dissipation
drives such a system to a fixed point for which the eccentricities
cease to exhibit secular oscillations and the apsidal lines of the two
planets precess at the same rate (\citealt{wugoldreich02};
\citealt{mardling07}). The apsidal precession rate of the inner planet
depends on a number of effects: general relativity; secular forcing by
the outer planet; the rotational bulge of the inner planet (assumed to
be in synchronous rotation with its orbit); and finally the tidally
distorted figure of the inner planet.  The latter two effects depend
on $k_2$.  By careful measurement of the Keplerian orbital elements
via high-quality transit and Doppler observations, the total
precession rate can be directly evaluated, as can all of the
individual contributions to the inner planet's precession rate --- all
except those contributions that depend on $k_2$, whose value can then
be backed out.

To date this method has been applied to the HAT-P-13 b-c system,
revealing that the inner hot Jupiter has $0.265 <k_2< 0.379$ 
and $M_{\rm core} < 27 M_{\oplus}$ \citep{Batyginetal09,Krammetal12}.
Extension of the method to close-in super-Earths should be
possible, particularly with TESS \citep{Rickeretal10}.

%
%

\section*{\small{Acknowledgments}}
\small{We thank Jacob Bean, Peter Bodenheimer, Josh Eisner, Jonathan Fortney, Uma Gorti, Richard
  Greenberg, Edwin Kite, Marc Kuchner, Eric Lopez, Geoff Marcy, Chris Ormel, Margaret Pan, Ilaria Pascucci, Erik
  Petigura, Sean Raymond, Damien S\'egransan, Angie Wolfgang, and Andrew Youdin for helpful discussions. An anonymous referee provided a detailed report that led to substantive changes in this paper. EC
  acknowledges support from NSF grant AST-0909210. GL acknowledges
  support from NASA grant NNX11A145A. This study was fostered by the
  Bay Area Consortium for Exoplanet Science (BACES), whose members include NASA
  Ames, Berkeley, UC Santa Cruz, and the SETI Institute.}

%
%

\bibliographystyle{aa}
\bibliography{insitu.bib}

\end{document}